\documentstyle[aps,12pt]{revtex}

\begin{document}
\title{Dilaton black holes in grand canonical ensemble near the extreme state}
\author{O.B.Zaslavskii}
\address{Department of Physics, Kharkov State University, Svobody Sq.4, Kharkov\\
310077, Ukraine\\
E-mail: oleg.b.zaslavskii@univer.kharkov.ua}
\maketitle

\begin{abstract}
Dilaton black holes with a pure electric charge are considered in the frame
work of a grand canonical ensemble near the extreme state. It is shown that
there exists such a subset of boundary data that the Hawking temperature
smoothly goes to zero to an infinite value of a horizon radius but the
horizon area and entropy are finite and differ from zero. In string theory
the existence of a horizon in the extreme limit is due to the finiteness of
a system only.
\end{abstract}

\draft
\pacs{04.70Dy, 97.60Lf, 04.20 Jb}




The study of the nature of the extreme state is one of the ''hottest'' areas
in current researches on black hole physics. Especially this concerns the
possibility of their thermodynamic description and behavior of entropy and
temperature near such a state. In this respect ''ordinary'' black holes, for
example Reissner-Nordstr\"{o}m spacetimes (RN), and dilaton ones are in some
sense complementary to each other. For a RN black hole with mass $m$ and
charge $q$ the Hawking temperature $T_{H}\propto \sqrt{m^{2}-q^{2}}$goes to
zero smoothly when the ratio $r_{+}/r\rightarrow 1$ ($r_{+}$ $=m+\sqrt{%
m^{2}-q^{2}}$ is the radius of the event horizon, $r_{-}=m-\sqrt{m^{2}-q^{2}}
$ is that of an inner one). In so doing, the entropy $S=A/4$ ($A$ is the
horizon area) remains finite nonzero quantity $\pi r_{+}^{2}$. On the other
hand, for dilaton black holes \cite{[1]}, \cite{[2]}, \cite{[3]} $%
T_{H}=(4\pi r_{+})^{-1},$ while $S=\pi r_{+}^{2}(1-r_{-/}r_{+})$ where $%
r_{+} $ and $r_{-}$ are parameters of the metric whose explicit form is
listed below. Thus, for RN black holes there exists the limit $r_{+/}r_{-}$ $%
\rightarrow 1$ in which $T_{H}\rightarrow 0,$ $S\neq 0$ and for dilaton
black holes there exists the limit in which $T_{H}\neq 0,$ $S\rightarrow 0$
(the extreme state).

Meanwhile, it has been shown recently \cite{zasl96}, \cite{zasl97} that
careful treatment in the framework of gravitational thermodynamics which
takes into account properly the finiteness of a system size displays
qualitatively new features in the geometry and thermodynamics of RN black
holes near the extreme state. In particular, it turned out that the extreme
state can be achieved at {\it finite temperature} $T$ that determines
properties of a canonical and grand canonical ensemble and differs from $%
T_{H}$ by blueshifting factor \cite{york86}. It is natural to pose the same
problem for dilaton black holes and examine the possible role for them of
the finiteness of a system.

In the present paper I show that this role is highly nontrivial. On the
basis of treatment in an infinite space it was believed that in the extreme
limit the horizon inevitably becomes singular with zero surface area \cite
{[3]}. However, it turns out that there exist such configurations that the
limit $r_{+}/r_{-}\rightarrow 1$ can be achieved with the a {\it regular}
horizon and {\it finite entropy.} It is essential that this result is
obtained in a model-independent way in view of primary principles of
gravitational thermodynamics. It occurs even in the simplest case of a pure
electrically charged holes to which we restrict ourselves in this issue.
(Therefore, this result should not be confused with those for extreme
dilaton black holes in supersymmetric theories \cite{[4]} where the property 
$S\neq 0$ for them is due to the simultaneous existence of both electrical
and magnetic charges.) As far as temperature is concerned, it turns out that 
$T_{H}\rightarrow 0$ smoothly but $T$ remains finite.

Apart from this, I discuss briefly the example from Ref. \cite{[3]} (where
black holes with a pure magnetic charge are considered) for the metric in a
string theory. Here gravitational thermodynamics predicts that in addition
to a bottomless hole without a horizon described in \cite{[3]} a black hole
also may exist in the limit $r_{+}=r_{-}.$

Both results rely strongly on the finite size of a system which enters as
one of boundary data in their complete set in the grand canonical approach
which from methodical viewpoint generalized the approach of Ref. \cite{[5]}
developed for black holes in the Einstein-Maxwell theory to the case of
dilaton ones.

It follows from \cite{[1]}-\cite{[3]} that the Euclidean metric of dilaton
black holes with a pure electric charge (or with a pure magnetic one) reads 
\begin{equation}
ds^{2}=(1-r_{+}/r)d\tau ^{2}+(1-r_{+}/r)^{-1}dr^{2}+R^{2}d\omega ^{2}
\label{(1)}
\end{equation}
where $d\omega ^{2}$ is the metric on a sphere of an unit radius. For this
spacetime 
\begin{equation}
R^{2}=r^{2}(1-r_{-}/r),\text{ }e^{2\varphi }=e^{2\varphi _{0}}(1-r_{-}/r),%
\text{ }F=Q/r^{2}dt\wedge dr,\text{ }Q^{2}=\frac{r_{+}r_{-}}{2}e^{2\varphi
_{0}}  \label{(2)}
\end{equation}
Here $F$ is electromagnetic field, $\varphi $ is a dilaton, the constant $%
\varphi _{0}$ for an infinite space coincides with $\varphi (r=\infty ).$ To
avoid possible confusion with the sign of $\varphi ,$one should bear in mind
that we discuss the case of a black hole with a pure electric charge whereas
in Ref.\cite{[3]} a charge is pure magnetic. Henceforth, a black hole is
supposed to be enclosed in a cavity. Then the state of a system in a grand
canonical ensemble is determined by boundary data \cite{[5]} whose complete
set contains in our case $\beta ,$ $\phi _{B},$ $\varphi _{B},$ $R_{B}.$
Here $\beta $ is the inverse Tolman temperature on a boundary, $\phi _{B}$
is the difference of potentials between a horizon and boundary, $\varphi
_{B} $ is a dilaton value on a boundary. The key moment consists in that
just $R_{B}$ determines the metric induced on a boundary surface and its
area $4\pi R_{B}^{2}$ and for this reason $R_{B}$ but not $r_{B}$ must enter
the set of boundary data for a thermal ensemble. (A radial coordinate $r$ is
in fact an auxiliary variable in terms of which the surface area should be
expressed. One could, in principle, rescale $r$ and obtain for it another
boundary value with the same $R_{B}.$)

Following primary principles of gravitational thermodynamics \cite{york86}
which takes into account properly space inhomogeneity of gravitating systems
one should write down relevant quantities (the inverse temperature, surface
area and the value of dilaton field on our case) as functions of coordinates
and equate them with their boundary values. As a result, we obtain equations 
\begin{equation}
\beta =4\pi r_{+}(1-r_{+}/r_{B})^{1/2}  \label{(3)}
\end{equation}
\begin{equation}
R_{B}=r_{B}(1-r_{-}/r_{B})^{1/2}  \label{(4)}
\end{equation}
\begin{equation}
\phi =Q(r_{+}^{-1}-r_{B}^{-1})(1-r_{+}/r_{B})^{-1/2}  \label{(5)}
\end{equation}
Here eq. \ref{(3)} takes into account that $\beta $ is blueshifted according
to the Tolman formula with the Hawking temperature $T_{H}=(4\pi r_{+})^{-1}.$
Eq. \ref{(5)} plays the similar role for the potential (cf. with
Einstein-Maxwell black holes \cite{[5]}). A charge $Q$ in terms of boundary
data $\varphi _{B}$ instead of $\varphi _{0}$ equals 
\begin{equation}
Q^{2}=\frac{r_{+}r_{-}}{2}e^{2\varphi _{B}}(1-r_{-}/r_{B})^{-1}  \label{(6)}
\end{equation}

After substitution of eq.\ref{(6)} into eq.\ref{(5)} we have three equations
for three variables $r_{B},$ $r_{+}$ and $r_{-}$ in terms of boundary data $%
\beta ,R_{B},$ $\varphi _{B},\phi _{B}.$ As is shown below, $\phi _{B}$ and $%
\varphi _{B}$ enter the set of boundary data in a single combination, so we
have three equations for three variables. (In the simplest case of
Schwarzschild black hole $R=r$ and one would have only one eq.\ref{(3)} from
which the horizon radius is to be defined as $r_{+}=r_{+}(r_{B},\beta )$ 
\cite{york86}.) Once $r_{B},$ $r_{+}$ and $r_{-\text{ }}$ are determined in
terms of boundary data one can find, for example, the horizon area in terms
of these data.

Now we will show that among all possible boundary data there exists such a
subset with {\it finite} $\beta $, $\phi _{B}$ and $R_{B}$ that the ratio $%
r_{+}/r_{-}\rightarrow 1$ in such a way that the following relations hold 
\begin{equation}
r_{+}/r_{B}\rightarrow 1,\text{ }r_{+},r_{-},r_{B}\rightarrow \infty ,\text{ 
}T_{H}\rightarrow 0,\text{ }A\neq 0  \label{(7)}
\end{equation}

(I recall that $r_{B}$ in contrast to $R_{B}$ is not a fixed parameter but a
variable to be determined from eqs. \ref{(3)}-\ref{(6)}.)

It is convenient to introduce dimensionless variables $x=r_{+}/r_{B},$ $%
y=r_{-}/r_{B},$ $\sigma =\beta /4\pi R_{B}.$ Then eqs. \ref{(3)}-\ref{(6)}
take the form 
\begin{equation}
\sigma =x(1-x)^{1/2}z^{-1},\text{ }z=(1-y)^{1/2},\text{ }\alpha \equiv \phi
_{B}e^{-\varphi _{B}}\sqrt{2}=[y(1-x)/x(1-y)]^{1/2}  \label{(8)}
\end{equation}

Note that either $Q$ or $\varphi _{0}$ do not enter \ref{(8)} explicitly as
they are eliminated according to $\ref{(5)},$ \ref{(6)}.

It follows from \ref{(8)} that 
\begin{equation}
\alpha ^{2}=\sigma ^{2}x^{-3}+1-x^{-1}  \label{(9)}
\end{equation}

Let $\alpha =\sigma .$ Then eq. \ref{(9)} has the root $x=1.$ In so doing, $%
y=1,$ $z=0$ according to $\ref{(8)}.$ If $x=1-\varepsilon $ and $y=1-\delta $
where $\varepsilon ,\delta <<1$ eq. $(8)$ shows the law according to which
the point ($x,y)$ approaches the limit under discussion: $\varepsilon
/\delta =\alpha ^{2}=\sigma ^{2}.$ Returning to dimension variables we see
that conditions \ref{(7)} are indeed satisfied.

Solutions found above can be also obtained with the help of the Euclidean
action. The crucial role in this approach is played by the finiteness of a
system as was first demonstrated in \cite{york86} for Schwarzschild black
holes. In particular, there exists such a range of boundary data that
Euclidean action not only has a local minimum as a function of a horizon
radius but also this minimum is global, so $I<0$ (it is assumed that $I=0$
for a hot flat space). It means that for a corresponding set of boundary
data a black hole spacetime is a favorable phase, so placing a system into a
box with spherical walls can stabilize such a state. This is intimately
connected with the space inhomogeneity of self-gravitating systems which
reveals itself in the crucial role of boundary conditions (first of all, it
makes a canonical or grand canonical ensemble for black holes well defined
and, in particular, leads to the possibility of a positive heat capacity for
a Schwarzschild black hole which was believed earlier to be only unstable).
Below we will see that allowance for the finiteness of a system for dilaton
holes leads to the possibility of a stable phase for the problem under
discussion as well.

The Euclidean action reads 
\begin{equation}
I=\beta E-S-\beta \phi _{B}q  \label{(10)}
\end{equation}

Here the constant $q$ appears in the analog of the Gauss law which can be
obtained by integrating the field equation $[F_{01}R^{2}\exp (-2\varphi
_{0})],r=0.$ This constant $q=Q\exp (-2\varphi _{0})$ differs from $Q$ due
to coupling between the electromagnetic and electric fields. The entropy $%
S=A/4,$ $E$ is the energy. The energy density $\varepsilon =(k-k_{0})/8\pi $ 
\cite{[6]}, \cite{[7]} where $k$ is the extrinsic curvature of the boundary
in three-dimensional spatial metric of slices $\tau =const.$ The constant $%
k_{0}$ is chosen to make $I=0$ for a hot flat space. Calculating $k$ and
using dimensionless variables $x=r_{+}/r$ and $y=r_{-}/r$ one obtains the
expression 
\begin{equation}
J\equiv I/4\pi R_{B}^{2}=\sigma [1-(1-\frac{y}{2})(1-x)^{1/2}(1-y)^{-1/2}]-%
\frac{(x^{2}-xy)}{4(1-y)}-\frac{\sigma \alpha }{2}(xy)^{1/2}  \label{(11)}
\end{equation}

It is convenient to introduce new variables $\eta =(1-x)^{1/2}(1-y)^{-1/2}$
and $\delta =1-y.$ Then for $\alpha =\sigma $ we get dropping all terms of
the third order and higher in $\delta :$%
\begin{equation}
J=2(\eta -\sigma )^{2}+\delta [(\eta -\sigma )^{2}+\eta ^{2}(\sigma
^{2}-\eta ^{2})]+\frac{\sigma ^{2}\delta ^{2}(1-\eta ^{2})^{2}}{4}
\label{(12)}
\end{equation}

It follows from \ref{(12)} that $\eta =\sigma ,$ $\delta =0$ ($y=1$) is
indeed a local extremum. If $\sigma ^{2}<\sqrt{2}-1$ this point is a
minimum. If, apart from this, $\sigma ^{2}<\frac{3}{2}-\sqrt{2}$ then $J<0,$
so this minimum is global, i.e. a black hole is thermodynamically favorable
phase as compared to a hot flat space. Thus, there exists a finite range of $%
\sigma $ within which the found solution is stable either locally or
globally.

Now we will find the form of the metric in the state under discussion. As $%
r_{+}<r<r_{B}$ and $r_{+}/r_{B}\rightarrow 1$ the coordinate $r$ becomes
singular not only near the horizon but for the {\it whole }Euclidean
manifold. It is convenient to use instead of $r$ the proper distance $l$
from the horizon $l.$ In the limit under consideration $r\rightarrow r_{B}$
and $r-r_{+}=l^{2}/4r_{B}$ where $r_{B\,}\rightarrow \infty .$ Let the
Euclidean time be normalized according to $0\leq \tau \leq 2\pi .$ Then
expressing $r$ in terms of $l$ we get after simple calculations with eqs. 
\ref{(3)}, \ref{(8)} taken into account: 
\begin{equation}
ds^{2}=d\tau ^{2}l^{2}+dl^{2}+[R_{B}^{2}+\frac{(l^{2}-l_{B}^{2})}{4}]d\omega
^{2}  \label{(13)}
\end{equation}

It follows from the Tolman formula that $\beta =2\pi l_{B}$ ($l_{B}=2\sigma
R_{B})$ and we see that the horizon area $A=4\pi R_{B}^{2}(1-\sigma ^{2})$
is a finite nonzero quantity. (Note that just $R_{B}$ but not $r_{B}$ enters
the final expression for $A$ in accordance with what was said above about
relation between these quantities.)

The metric \ref{(13)} can be also rewritten in the Schwarzschild-like form 
\begin{equation}
ds^{2}=4(\rho ^{2}-\rho _{0}^{2})d\tau ^{2}+4d\rho ^{2}(1-\rho _{0}^{2}/\rho
^{2})^{-1}+\rho ^{2}d\omega ^{2}  \label{(14)}
\end{equation}
where $\rho ^{2}=R_{B}^{2}+(l^{2}-l_{B}^{2})/4,$ $\rho
_{0}^{2}=R_{B}^{2}-l_{H}^{2}/4.$

Let us discuss now another case - metric in string theory that was
considered in Ref. \cite{[3]} with a pure magnetic charge. The relevant
metric $\tilde{g}_{\mu \nu }$ is obtained from the original one $g_{\mu \nu
} $ by conformal transformation with the factor $e^{2\varphi }.$ According
to \ref{(1)}, \ref{(2)} this factor in terms of boundary data equals 
\begin{equation}
\exp (2\varphi )=\exp (2\varphi _{B})(1-r_{-}/r_{B})^{-1}(1-r_{-}/r)
\end{equation}
Then omitting constant terms we obtain 
\begin{equation}
d\tilde{s}^{2}=(1-r_{+}/r)(1-r_{-}/r)^{-1}d\tau
^{2}+[(1-r_{+}/r)(1-r_{-}/r)]^{-1}dr^{2}+r^{2}d\omega ^{2}  \label{(15)}
\end{equation}
where $r$ takes finite values in the range $r_{+}\leq r\leq r_{B}.$ It was
pointed out in \cite{[3]} that in the extreme limit $r_{+}=r_{-}$ the metric
describes a bottomless hole ($l=\infty $ for any $r>r_{+})$ without an event
horizon.

However, it is worth paying attention that this is indeed the case only if
one put $r_{+}=r_{-}$ right from the beginning. If this limit is achieved
from a topological sector of black holes in the manner described above the
situation is qualitatively different. Now according to eq.\ref{(2)} and eq.%
\ref{(7)} the factor $\exp (2\varphi )$ is equal to $\exp (2\varphi
_{B})R^{2}/R_{B}^{2}$. It is finite and nonzero everywhere (whereas it
turned to the zero at the horizon $r=r_{+}=r_{-\text{ }}$ in the previous
case ), so the properties of the metric $\tilde{g}_{\mu \nu }$ in string
theory are qualitatively similar to whose of the original one $g_{\mu \nu }.$
In particular, the horizon does not disappear.

It is instructive to compare relationship between extreme and non-extreme
cases for RN and dilaton black holes in terms of geometrical
characteristics. In the first case the geometrical property which signals
about extremality consists in $l=\infty $ for any point with $r>r_{+}$
outside the horizon. Therefore, it turned out rather unexpectedly that in
the framework of the grand canonical ensemble the limiting transition can be
performed in such a way that $l$ remains finite when $r_{+}/r_{-}\rightarrow
1$ \cite{zasl96}, \cite{zasl97}. On the other hand, it{\it \ would seem}
obvious from results of \cite{[3]} that for dilaton black holes with a pure
electric (or pure magnetic) charge $A\rightarrow 0$ in the extreme limit $%
r_{+}/r_{-}=1.$ None the less, as it follows from the results of the present
paper, for the particular class of boundary data this quantity tends to the
finite nonzero limit.

Thus, in both cases careful treatment in terms of thermodynamics shows that
the appropriate choice of boundary data affects geometry and even topology
crucially. It is worth stressing that the key point here is the finiteness
of a system. Otherwise the solutions considered above would have been lost.
The situation is especially pronounced in the example from string theory
when the possibility of the existence of a horizon is due to the finiteness
of a system entirely.

The results obtained in this paper can be extended to the case of more
general coupling between dilaton and electromagnetic field in a
straightforward manner. What is more the mechanism by means of which
thermodynamic and geometrical properties are overlapped near the extreme
state for spacetimes similar to \ref{(1)} is insensitive to subtle details
of a field theory within which a general form \ref{(1)} is obtained.


%
%

%
%

\end{document}